\documentclass{PoS}

\title{WKB - type approximations in the theory of vacuum particle creation in strong fields}

\ShortTitle{WKB - type approximations}

\author{\speaker{S.A. Smolyansky}, V.V. Dmitriev, A.D. Panferov and \fbox{A.V. Prozorkevich}\\
Saratov State University, Saratov, Russia\\
E-mail: \email{smol@sgu.ru}, \email{dmitrievv@gmail.com},
 \email{panferovad@info.sgu.ru}}

\author{D. Blaschke, L. Juchnowski\\
Institute for Theoretical Physics, University of Wroclaw, 50-204;
Wroclaw, Poland \\
E-mail: \email{blaschke@ift.uni.wroc.pl}, \email{lukasz.juchnowski@ift.uni.wroc.pl}}

\abstract{Within the theory of vacuum creation of an $e^{+}e^{-}$ -
plasma in the strong electric fields acting in the
focal spot of counter-propagating laser beams we compare
predictions on the basis of different WKB-type approximations with 
results obtained in the framework of a strict kinetic approach. 
Such a comparison demonstrates a considerable divergence results. 
We analyse some reasoning for this observation and conclude that 
WKB-type approximations have an insufficient foundation in the 
framework of QED in strong nonstationary fields. 
The results obtained in this work on the basis of the kinetic approach 
are most optimistic for the observation of an $e^{+}e^{-}$ -
plasma in the range of optical and x-ray laser facilities. 
We discuss also the influence of unphysical
features of non-adiabatic field models on the reliability of
predictions of the kinetic theory.}

\FullConference{XXII International Baldin Seminar on High Energy Physics Problems,\\
        15-20 September 2014\\
        JINR, Dubna, Russia}

\begin{document}

\section{Introduction}

As it is known, the first predictions of the vacuum creation of an
$e^{+}e^{-}$ - plasma (EPP) under the action of a constant electric
field appeared in the discussion of Klein's paradox \cite{0} based 
on the interpretation as a tunnel effect \cite{1,2}. 
Recently, the tunneling mechanism has been applied to predict chiral 
quasipartcile pair creation in graphene \cite{graphene}. 
With the development of high-intensity lasers the idea was raised 
to use strong laser fields for the verification of the 
"Schwinger effect" \cite{schwinger}. 
At the present time there are realistic projects for future
experiments of such kind at high-power laser facilities 
(see, e.g., Ref. \cite{3}). 
These perspectives and the success of the Sauter-Heisenberg-Euler
prediction have excited hope that a similar approach could be used
in the case of "laser" fields. 
This resulted in the classical work by Brezin and Itzykson \cite{4} 
and a series of works by V.S. Popov \cite{5} based on different versions 
of the WKB approximation in the QED of strong nonstationary fields.
In this context let us mention also the work \cite{6}. 
On the other hand, there were also doubts raised in the validity of
WKB-type approaches to the case of fastly alternating fields expressed,
e.g., in \cite{7}.

The idea of the present work is to verify the correctness of the results of
different WKB-type approaches for fast "laser" fields by using the strong 
kinetic equation (KE) approach \cite{8} as the basis. 
It is important that for a comparison we use the same nonadiabatic field
model of a periodical signal as in the works \cite{4,5,6}. 
Such a comparison shows that one can speak at best of a qualitative
similarity in the asymptotic regions of low and high frequency
of the electric field. 
Differences become very large in the region of intermediate frequencies. 
We analyse some details of this picture. 
On the other hand, the usage of the unphysical model of a nonadiabatic 
electric field in the kinetic approach can lead to a considerable distortion 
of the EPP creation pattern.

\section{Kinetic approach}

Let us write the system of KE's \cite{9} as a system of
ordinary differential equations (ODE)
\begin{equation}
\dot{f}=\frac{1}{2}\lambda u,~~~\dot{u}=\lambda (1-2f)-2\varepsilon v,~~~
\dot{v}=2\varepsilon u,  \label{1}
\end{equation}
that is equivalent to the KE in the integro-differential form \cite{8}. 
In Eq. (\ref{1}) $f(\vec{p},t)$ is the distribution function, 
$\varepsilon (\vec{p},t)=
\sqrt{ \varepsilon _{\bot }^{2}+(p_{3}-eA(t))^{2}}$
is the quasienergy of a charge particle in a time dependent
electric field of the linear polarization 
$A^{\mu }(t)=\left(0,0,0,A^{3}=A(t)\right) ,~~~
\varepsilon _{\bot }=\sqrt{m^{2}+p_{\bot }^{2}}$ is the transverse energy and
$\lambda (\vec{p},t)=eE(t)\varepsilon _{\bot}/\varepsilon ^{2}$ 
is the amplitude of the vacuum excitation, $E(t)=-\dot{A}(t)$.

We solve this system of ODE's numerically with zero initial conditions 
$f_{0}=u_{0}=v_{0}=0$ at $t=0$ for the nonadiabatic field model
\begin{equation}
E(t)=E_{0}\sin \xi (t),~~~A(t)=(E_{0}/\omega )\cos \xi (t),
\label{2}
\end{equation}
where $\xi (t)=\omega t+\varphi ,$ $\varphi $ is an initial phase. 
We are forced to use this unphysical field model as a basic one in the 
wake of the works of the WKB direction. 
There are two limiting cases:
1) $\varphi ^{(1)}=0$ (the field strength starts from
zero, $E(0)=0$, but with nonzero derivative, $\dot{E}(0)\neq 0$, and 
$A(0)\neq 0$) and 
2) $\varphi ^{(2)}=\pi /2$ (sharp switching on of the field, 
$E(0)=E_{0}\neq 0$). 
The WKB-type approaches are insensitive to choice of the phase. 
However, the ODE\ system (\ref{1}) takes these two initial conditions as 
quite different.
Keeping in mind some matching with the WKB\ results we will consider
the first case ($\varphi ^{(1)}=0$) only because of the second
variant of switching on leads to explosive EPP creation (in the
framework of the kinetic approach). On the other hand, both variants
of switching on the field are unphysical and generate an incorrect
statement of the initial value problem for the ODE system (\ref{1}).

According to the works \cite{4,5,6} of the WKB\ direction the out-state in
the field model (\ref{2}) with $\varphi ^{(1)}=0$ is defined in the
discrete set of the time points $t_{n}=(T/2)n,~~~n=1,2,...$, i.e.,
$f_{\rm out}=f_{n}=f(t_{n})$, with $T$ being the period of the field (\ref{2}). 
The corresponding pair number density of the residual EPP will be 
\begin{equation}
n_{n}=2\int \frac{d^{3}p}{(2\pi )^{3}}f_{n}.  
\label{3}
\end{equation}
The production rate $w_{n}=2f_{n}/nT$ shall be related to the volume 
$\lambda ^{3}$, i.e. $w_{n}/\lambda ^{3}$, where $\lambda $ is the wavelength,
since $T=\lambda $. 
We are interested in the initial linear region of the effect ($n=1,2$).

The function $w_{n}$ is considered as a function of two dimensionless 
field parameters $\eta =E_{0}/E_{c}$ ($E_{c}=m^{2}/|e|$ is the critical field) 
and the adiabaticity parameter $\gamma =E_{c}\omega /E_{0}m$, which allows 
to separate the regions of the tunneling ($\gamma <<1$)
and multiphoton ($\gamma >>1$) mechanisms of the vacuum creation. 
For $\eta =const$, the parameter $\gamma $ plays the role of the angular frequency. 
Then, for example, $\lambda =E_{c}/(2\pi mE_{0}\gamma )$.

\section{Comparison}

For definiteness we cite the concrete formulas that are the results
of the WKB\ analysis.

In the classical work \cite{4} the pair production rate per unit
volume was obtained for charged bosons with the electron charge
and mass ($\alpha =e^{2}/4\pi $)
\begin{equation}
w_{BI}=\frac{\alpha E_{0}^{2}}{2\pi }\frac{1}{g(\gamma )+\gamma
g^{\prime }(\gamma )/2}\exp \left[ -\pi \left( E_{c}/E_{0}\right)
g(\gamma )\right] , \label{4}
\end{equation}
where the smooth function $g(\gamma )$ is
\begin{equation}
g(\gamma )=\frac{4}{\pi }\int_{0}^{1}dy\left(\frac{1-y^{2}}{1+\gamma ^{2}y^{2}}\right) ^{1/2}.  
\label{5}
\end{equation}
The asymptotics for small and large $\gamma $ are given by
\begin{equation}
w_{BI}\simeq \frac{\alpha E^{2}}{2\pi }\exp \left[ -\pi
E_{c}/E_{0}\right] ,~~~\gamma <<1;  \label{6}
\end{equation}
and
\begin{equation}
w_{BI}\simeq \frac{\alpha E^{2}}{8}\left( \frac{eE_{0}}{2m\omega
}\right) ^{4m/\omega },~~~\gamma >>1~.  
\label{7}
\end{equation}
Different results are given in the series of V.S. Popov's works. 
We will follow the work \cite{10} where the pre-exponential factor is fixed
\begin{equation}
w_{P}=\frac{m^{4}}{\pi ^{3}\sqrt{a_{1}}}\left(
\frac{E_{0}}{E_{c}}\right) ^{5/2}\exp \left[ -\pi
(E_{0}/E_{c})g(\gamma )\right] ,  
\label{8}
\end{equation}
where $g(\gamma )$ is defined by Eq.~(\ref{5}), and has the asymptotics
\begin{equation}
g(\gamma )=\left\{
\begin{array}{c}
1-\gamma ^{2}/8,~~~\gamma <<1; \\
4(\pi \gamma )^{-1}(\ln \gamma +a_{1}),~~~\gamma >>1
\end{array}
\right.  
\label{9}
\end{equation}
and $a_{1}=1,47$.

For comparison, we remind also Schwinger's classical formulas for electrons
\begin{equation}
w_{(1/2)}=\frac{\alpha E_{0}^{2}}{\pi ^{2}}\exp \left( -\pi
E_{c}/E_{0}\right)  \label{10}
\end{equation}
and $w_{(0)}=w_{(1/2)}/2$ for bosons (the fermionic dominance).

A reasonable expectation would be that the WKB approximate
solutions must correspond to the Schwinger result (\ref{10}) in the
limit $\gamma \rightarrow 0$ (i.e. $\omega \rightarrow 0$ for $E_0 =
const$). 
However, the Eqs.~(\ref{4}) and (\ref{8}) do not obey to this correspondence. 
As far as the exponential factors are equal at $\gamma \rightarrow 0$  under 
Eq.~(\ref{9}), we will compare the pre-factors in this limit
\begin{equation}
1)~ \frac{w_{BI}}{w_{(0)}} = \pi, ~~~ 
2)~ \frac{w_{P}}{w_{(1/2)}} =
\frac{4}{\sqrt{a_{1}}}\left( \frac{E_{0}}{E_{c}}\right) ^{1/2}, ~~~
3)~ \frac{w_{BI}}{w_{(1/2)}} = \pi/2 >1. 
\label{11}
\end{equation}
The last relation indicates a violation of the fermionic dominance
rule in the approach \cite{4}. 
Nontriviality of the relations (\ref{11}) hints to problems in the construction 
of the low frequency limit of the WKB type approximations in general.

\begin{figure}[!htb]
\begin{minipage}{0.5\textwidth}
\includegraphics[width=1\textwidth]{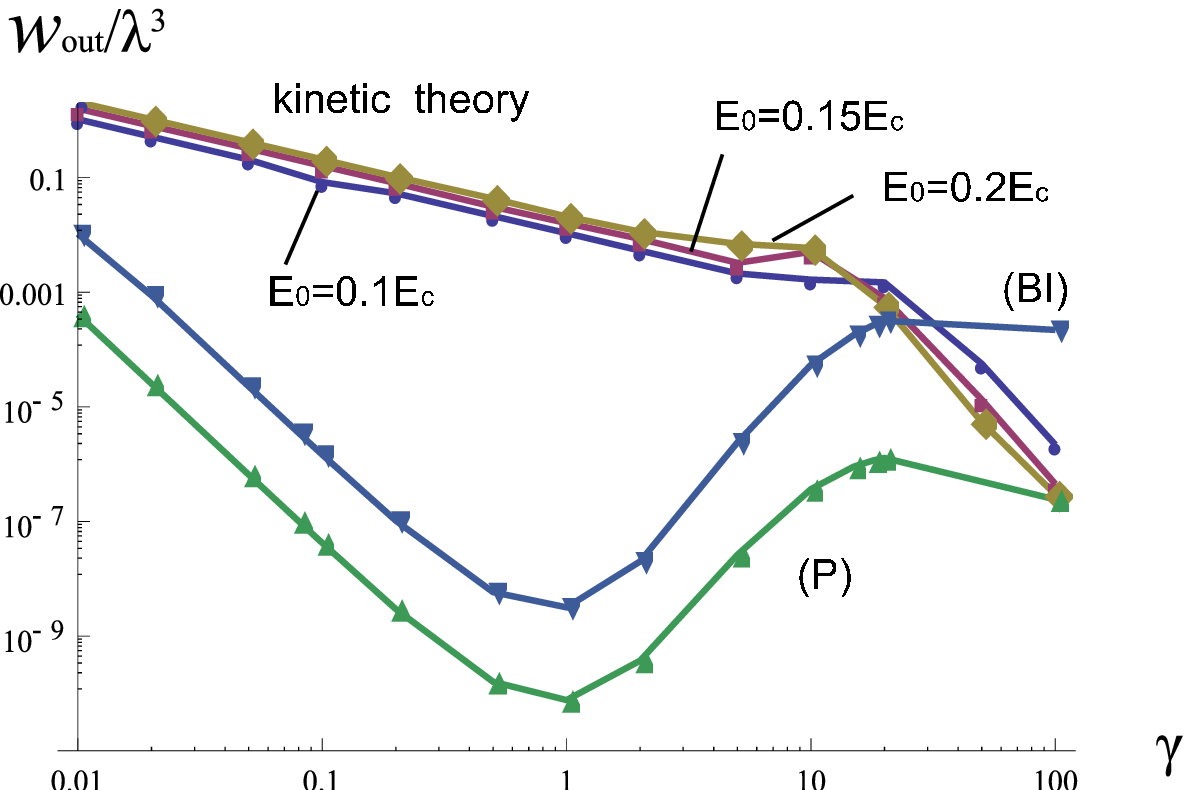}
\end{minipage}\hfill
\begin{minipage}{0.5\textwidth}
\includegraphics[width=1\textwidth]{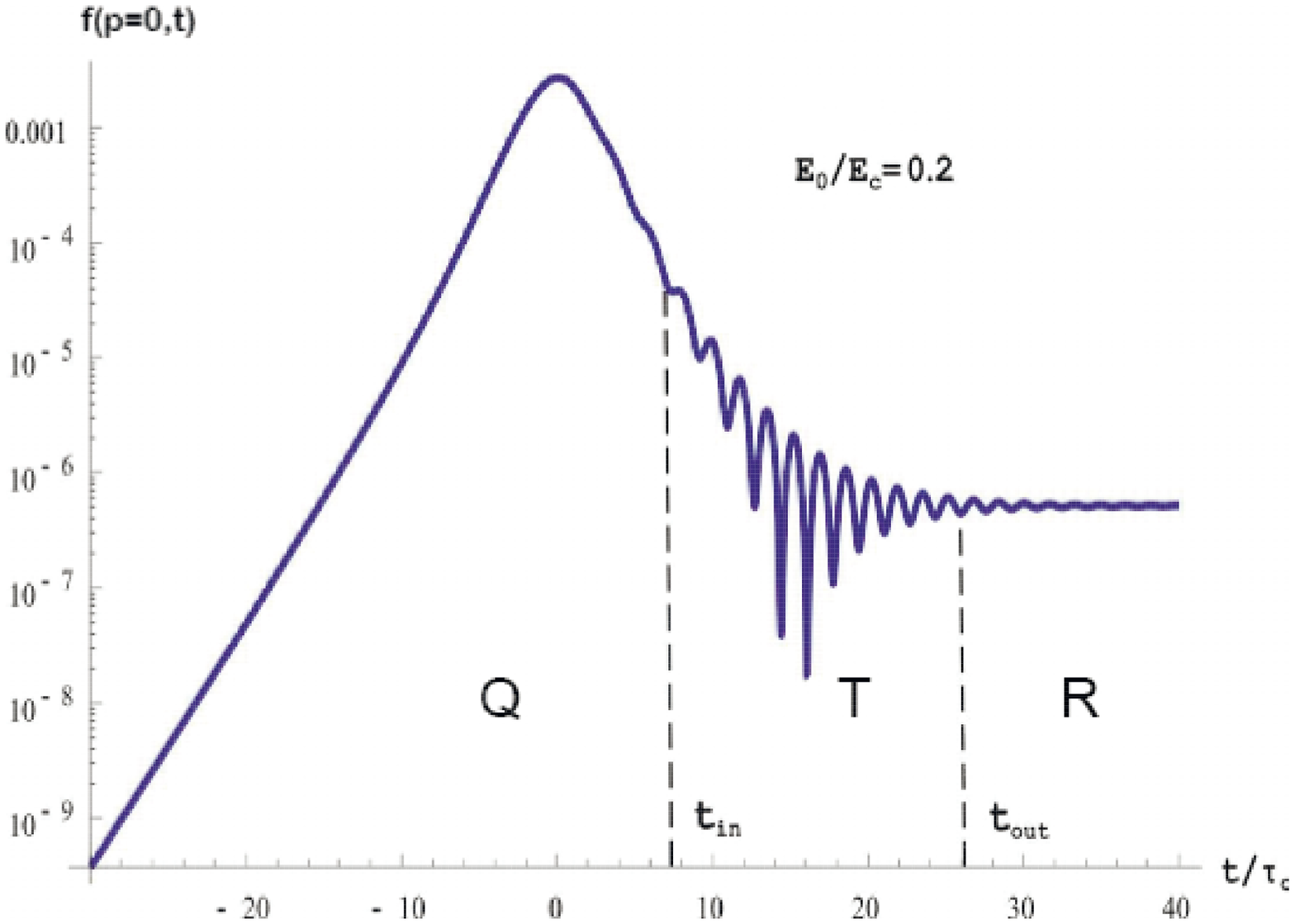}
\end{minipage}
\caption{\textbf{Left panel:} Comparison the EPP production rates obtained
in the framework of the WKB type approximations of Brezin-Itzykson
(BI) and Popov (P) with the result of the kinetic approach (three curves
for $E_{0}=0.1E_{c}$, $E_{0}=0.15E_{c}$ and $E_{0}=0.2E_{c}$ resp.)
demonstrates the stability of predictions within the KE approach.
\textbf{Right panel:} Evolution of the EPP distribution function
$f(\vec{p}=0,t)$ under action of the Eckart field. One distinguishes 
(Q)uasiparticle, (R)esidual and (T)ransient regions of EPP production. } 
\label{fig}
\end{figure}

A comparison of results of the kinetic theory (Eqs.~(\ref{1}),(\ref{3}))
with Eqs.~(\ref{4}) - (\ref{7}) (the Brezin - Itzykson approach) 
and Eqs.~(\ref{8}), (\ref{9}) (the imaginary time method \cite{10}) are
presented in Fig.~\ref{fig} (the pair production rates (\ref{4})
and (\ref{8}) were brought to the focal spot volume). 
This figure demonstrates the tendency to some qualitative agreement at 
$\gamma \rightarrow 0$. 
However, the difference becomes appreciable ($\sim 10^{3}$) already for 
$\gamma =0.01$. 
This difference grows with increasing $\gamma $. 
The violation of the fermionic dominance rule when comparing the Brezin-Itzykson 
and Popov approaches is observed for any $\gamma > 0$.

Comparison of the results of the kinetic theory and WKB approaches
in the left panel of Fig.~\ref{fig} for $E_{0}=0,1E_{c}$ and two other
field strengths ($E_{0}=0.15 E_{c}$ and $E_{0}=0.2 E_{c}$) demonstrates 
the stability of predictions of the kinetic theory.

The general reason for such a remarkable difference is the extreme 
sensibility of the phenomena of the EPP vacuum creation to details of temporal
behaviour (or the frequency spectrum) of an external field pulse \cite{11,otto,mb32}. 
As an example we show the evolution of the EPP distribution function 
$f(\vec{p}=0,t)$ generated by the smooth Sauter field 
$E(t)=E_{0}\cosh^{-2}(t/T)$ in the right panel of Fig.~\ref{fig}. 
Even in such simple field a very complicated structure is formed in the 
transient region which precedes the formation of the residual EPP in the 
out-state. 
Such kind of details are smoothed at the WKB\ roughening. 
It is not accidental that the asymptotic analysis of the ODE system
(\ref{1}) was effective for describing the quasiparticle stage
of the evolution only \cite{12}.

On the other hand, the exact kinetic description leads to some
phantom fragment of EPP distribution in the moment when the nonadiabatic field 
(\ref{2}) is switched on. 
These phantom distributions disappear for more realistic field models with soft
switching of the field. 
In addition, the nonadiabatic switching off at the moments $t_{n}$ is not 
corresponding to the usual understanding of the out-state because $E(t_{n})=0$ 
but $\dot{E}(t_{n})\neq 0$.
Within the framework of the KE description of EPP production the estimations 
in the field (\ref{2}) turn out overrated in comparison with more realistic 
field models. 
One can expect that the latter will lead to more realistic estimations 
of the pair production rate of EPP, especially, in the region $\gamma \sim 1$. 
We plan to investigate this in our subsequent work.

Finally, we would like to draw the attention also to the incomprehensible
concentration of all solutions in the region $\gamma \gg 1$, as shown in the left panel 
of Fig.~\ref{fig}.

\section{Conclusion}

Thus we have shown that there is insufficient foundation to trust 
estimations of the residual EPP production rate based on
different approximations of the WKB type. 
We come to the conclusion that the kinetic theory is now the unique 
adequate instrument for investigatiing the vacuum EPP creation for 
selections of reasonable physical models of the laser field.

We like to close with a reminiscence of prof. V.K. Lukyanov
about academician I.E. Tamm who confessed at the beginning of the
60-ies to his collaborators and aspirants at FIAN about his
unsuccessful attempts to adapt the WKB method to QED.

\subsection*{Acknowledgements}

S.A.S. grateful to M. Musakhanov for useful comments on the instanton 
interpretation of the formalism and to V.K. Lukyanov for remarks and support.
The work of D.B. was supported in part by the Polish Ministry of Science 
and Higher Education under grant No. 1009/S/IFT/14 and L.J. is grateful for 
support from the University of Wroclaw under internal grant No. 2467/M/IFT/14.

\end{document}